\documentclass{article}
\usepackage{arxiv}
\usepackage{hyperref}
\usepackage[T1]{fontenc}
\usepackage{xcolor} 
\usepackage{import}
\usepackage{graphicx}   
\usepackage{wrapfig}    
\usepackage{float}      
\usepackage{longtable}
\usepackage{booktabs}
\usepackage{enumitem}
\usepackage{subcaption}
\usepackage{amsmath} 
\usepackage{multirow}
\usepackage{multirow}
\usepackage{longtable} 
\usepackage{graphicx}
\begin{document}

\author{ 
	\href{https://orcid.org/0009-0005-0398-3115}{Joachim Andreassen} \\
	Department of Computer Science\\	
	Norwegian University of Science and Technology\\	
 Trondheim, Norway \\
	\texttt{joachan@stud.ntnu.no} \\
 \And%
\href{https://orcid.org/0009-0001-6072-4081}{Donn Morrison}\\
	Department of Computer Science\\
	Norwegian University of Science and Technology\\
	Trondheim, Norway\\
	\texttt{donn.morrison@ntnu.no} \\
}

\renewcommand{\shorttitle}{Discovery of Endianness and Instruction Size Characteristics in Binary Programs from Unknown Instruction Set Architectures}

\hypersetup{
pdftitle={Discovery of Endianness and Instruction Size Characteristics in Binary Programs from Unknown Instruction Set Architectures},
pdfsubject={q-bio.NC, q-bio.QM},
pdfauthor={Joachim Andreassen, Donn Morrison},
pdfkeywords={Reverse engineering, Unknown Instruction Set Architecture, Machine learning, Signal processing},
}

\title{Discovery of Endianness and Instruction Size Characteristics in Binary Programs from Unknown Instruction Set Architectures}

\maketitle              
\begin{abstract}

We study the problem of streamlining reverse engineering (RE) of binary programs from unknown instruction set architectures (ISA). We focus on two fundamental ISA characteristics to beginning the RE process: identification of endianness and whether the instruction width is a fixed or variable. For ISAs with a fixed instruction width, we also present methods for estimating the width. In addition to advancing research in software RE, our work can also be seen as a first step in hardware reverse engineering, because endianness and instruction format describe intrinsic characteristics of the underlying ISA.

We detail our efforts at feature engineering and perform experiments using a variety of machine learning models on two datasets of architectures using Leave-One-Group-Out-Cross-Validation to simulate conditions where the tested ISA is unknown during model training. We use bigram-based features for endianness detection and the autocorrelation function, commonly used in signal processing applications, for differentiation between fixed- and variable-width instruction sizes. A collection of classifiers from the machine learning library \texttt{scikit-learn} are used in the experiments to research these features. Initial results are promising, with accuracy of endianness detection at $99.4\%$, fixed- versus variable-width instruction size at $86.0\%$, and detection of fixed instruction sizes at $88.0\%$.


\end{abstract}

\section{Introduction}\label{new:10:introduction}
The emergence of IoT devices has increased the importance of understanding the workings of compiled binary files through reverse engineering (RE). Reverse engineering has applications in vulnerability research, extending support of legacy software and hardware, binary patching and translation, and digital forensics  \cite{towards_usable_cpu_architecture_detection}.
\\\\
Identifying the targeted binary file's instruction set architecture (ISA) is an essential first step in reverse engineering because it permits the reverse engineer to apply an appropriate disassembler to translate machine readable instructions into an assembly representation, and subsequently apply a decompiler that can yield high-level source code. Previous research has focused on this process by providing methods that classify ISAs reliably from a set of known ISAs \cite{automatic_classification_of_object_code_using_machine_learning,cpu_rec,towards_usable_cpu_architecture_detection,elisa_eliciting_isa_of_raw_binaries_for_fine_grained_code_and_data_separation}. However, identifying the ISA from binary files with an unknown or undocumented ISA has not been thoroughly explored previously. Proprietary ISAs with unavailable documentation and ISAs for custom virtual machines are common examples in this group of ISAs.
\\\\
This main aim of this work is to discover fundamental ISA characteristics from binary files where the ISA specification is either unknown, proprietary, or undocumented. This knowledge can be used to advance the reverse engineering process and generate documentation for unknown ISAs. As the understanding of the ISA becomes clearer, high-level program behavior such as control flow and call graph structure can be discovered \cite{pettersen2024graphdiscoverybinaryprograms}. 
\\\\
We study the following research questions:
\begin{enumerate}[leftmargin=1.1cm]
    \item[RQ1:] Can machine learning be used to detect intrinsic characteristics of unknown ISAs from binary programs?
    \item[RQ2:] Do the proposed approaches lead to reliable detection of ISA characteristics across a wide range of ISAs?
\end{enumerate}
\noindent
The main contribution of this paper is methods for discovering the following ISA characteristics:
\begin{enumerate}
    \item endianness of the ISA,
    \item fixed- versus variable-width instruction format of the ISA, and
    \item for fixed instruction width ISAs, an estimation of the width
\end{enumerate}
\noindent
The paper is structured as follows: Section \ref{new:20:background} presents background and related work. Section \ref{new:30:methodology} presents our methodology. Section \ref{new:40:results} presents the experimental setup and results. The results are then discussed in Section \ref{new:50:discussion}. A conclusion of the paper is provided in Section \ref{new:60:conclusion}.
\section{Background}\label{new:20:background}
A binary executable file consists of a series of executable instructions in a format understood by the CPU. When run by the CPU, a set of actions, primarily defined by a programmer, is executed. Binary executable files are generated from high-level code by a compiler that structures data into headers and segments. 
The headers contain information about the binary's properties and organization. The two segments central in this paper are the code and data segments. These include the executable code and global or static variables, respectively. The code segment is particularly central, as it contains a series of instructions from which the ISA characteristics of interest are detectable \cite{tis_elf}.
\\\\
The instruction set architecture (ISA) is an abstraction that specifies how CPU executes the instructions of binary programs. The specification of an ISA includes features such as endianness, instruction encoding and format, the number of physical registers, etc. \cite{arm_glossary_isa}.
\\\\
Endianness describes how multi-byte values and memory addresses are ordered. This paper focuses on the two most common endianness encodings: big and little. For big endianness, the most significant bit is stored in the lowest address and the least significant bit in the greatest address. The opposite is true for little endianness.
\\\\
Instruction size is the number of consecutive bits that define an instruction. The instruction size can be fixed or variable for a given ISA, and in some cases an ISA can support both fixed and variable formats (an example is the RISC-V ISA, which supports 32-bit width instructions with 16-bit extensions). Binary programs compiled for ISAs with a fixed instruction width contain only instructions of the specified size \cite{natural_language_processing_approach_for_isa_identification}.
\\\\



\subsection{Related Work}
Existing research in the field of ISA detection focuses on detecting the ISA from a predefined set of architectures. This differs from this paper's goal, which focuses on unknown ISAs that cannot be classified from a set of architectures, aiming to streamline reverse engineering of such architectures. However, these differences do not eliminate the relevance of previous research, where various approaches are usable in this research.
\\\\
A paper by Kairajärvi et al. \cite{isadetect} contributes in two ways. First, it provides the comprehensive \textit{IsaDetect} dataset with 66685 binary programs from 23 different architectures scraped from the Debian repositories. This is a balanced dataset, as each architecture contains a similar-sized sample. The dataset contains binary programs exclusively of either complete or code-only binary programs. These two versions of the dataset are in this paper referred to as \textit{IsaDetectFull} and \textit{IsaDetectCode}, respectively.
\\\\
Secondly, Kairajärvi et al. \cite{isadetect} explores state-of-the-art methods for detecting ISAs using the IsaDetect datasets. This involves a series of features used with machine learning. With this method, the paper can classify the ISA from a predefined set with an accuracy of $98\%$ with models trained and tested on the IsaDetectCode dataset.
\\\\
The CpuRec project by Granboulan \cite{cpu_rec} contribute a dataset and command-line tool for classifying ISAs. The tool computes the Kullback–Leibler divergence between a given binary executable file (the query) and each binary executable file in the dataset, where each file represents a different ISA. The query file is then classified as belonging to the ISA with the lowest Kullback–Leibler divergence.
\\\\
The \textit{CpuRec} \cite{cpu_rec} dataset is valuable for this research because it contains executable files from 77 different ISAs, each represented by a single code-only binary program. Although smaller than the IsaDetect dataset from \cite{isadetect}, it has broad coverage of diverse ISAs.
\\\\
Our use of endianness features builds on the work of Clemens \cite{automatic_classification_of_object_code_using_machine_learning}, who found that a histogram of bigrams yields information about endianness due to the contained information on byte-adjacency. However, the space of all possible bigrams yields a very large feature space ($256^2 = 65536$), which introduces a problem known as the \textit{curse of dimensionality}. The authors handle this problem by using a limited set of four bigrams: \texttt{0xfffe}, \texttt{0xfeff}, \texttt{0x0001} and \texttt{0x0100}.


\section{Methodology}\label{new:30:methodology}
This paper proposes machine learning to classify specific characteristics of ISAs in binary programs. We achieve this by engineering features that extract relevant information from the binary code. These features are then used to train and test machine learning models, enabling the identification of key ISA characteristics. To ensure unbiased testing, we use Leave-One-Group-Out Cross Validation (LOGOCV). LOGOCV involves training a separate model for each ISA in the dataset, where the specific ISA is left out of the training data and used only for testing. This approach allows us to simulate the real-world scenario of encountering a previously unseen ISA.
\\\\
This paper uses both the IsaDetectFull and CpuRec datasets. The IsaDetectFull dataset is used for endianness detection due to its large number of binary files and balance across big and little endianness. This dataset is favorable over the IsaDetectCode dataset as data affected by endianness also exists in parts of the binary file other than the code section (e.g., header information). The datasets and ISA characteristics are listed in Table \ref{new:tab:datasets_and_characteristics}.
\\

\footnotesize 

\begin{table}
\caption{Architectures from the IsaDetectFull and CpuRec datasets with endianness (E) and instruction size (IS) characteristics (in bits). Blank table elements mean that particular characteristic was not available for the dataset and thus was not used for model training.}
\label{new:tab:datasets_and_characteristics}
\centering\begin{tabular}{|l|l|c|c||l|l|c|c|}
\hline
\textbf{IsaDetectFull} & \textbf{CpuRec} & \textbf{E} & \textbf{IS} & \textbf{IsaDetectFull} & \textbf{CpuRec} & \textbf{E} & \textbf{IS} \\ \hline
~          & 6502               & LE & 8-32   & ~          & MMIX               & BE & 32    \\\hline
~          & 68HC08             & BE & 8-16   & ~          & MN10300            & LE & ~     \\\hline
~          & 68HC11             & BE & 8-40   & ~          & MSP430             & LE & ~     \\\hline
~          & 8051               & LE & 8-128  & ~          & Mico32             & BE & 32    \\\hline
arm64      & ARM64              & LE & 32     & ~          & MicroBlaze         & BI & 32    \\\hline 
~          & ARMeb              & BE & 32     & ~          & Moxie              & BI & 32-48 \\\hline 
armel      & ARMel              & LE & 32     & ~          & NDS32              & BI & 16-32 \\\hline 
armhf      & ARMhf              & LE & 32     & ~          & NIOS-II            & LE & 32    \\\hline 
~          & ARcompact          & LE & 16-32  & ~          & PDP-11             & LE & 16    \\\hline 
~          & AVR                & LE & 16-32  & ~          & PIC10              & LE & ~     \\\hline 
alpha      & Alpha              & LE & 32     & ~          & PIC16              & LE & ~     \\\hline 
~          & AxisCris           & LE & 16     & ~          & PIC18              & LE & ~     \\\hline 
~          & Blackfin           & LE & 16-32  & ~          & PIC24              & LE & 24    \\\hline 
~          & CLIPPER            & LE & 2-8    & ppc64      & PPCeb              & BE & ~     \\\hline 
~          & CUDA               & LE & 32     & ppc64el    & PPCel              & LE & ~     \\\hline 
~          & Cell-SPU           & BE & 32     & riscv64    & RISC-V             & LE & 32    \\\hline 
~          & CompactRISC        & LE & 16     & ~          & RL78               & LE & ~     \\\hline 
~          & Epiphany           & LE & 16-32  & ~          & ROMP               & BE & 8-32  \\\hline 
~          & FR30               & BE & 16     & ~          & RX                 & LE & ~     \\\hline 
~          & H8-300             & BE & 8-16   & s390x      & ~                  & BE & ~     \\\hline 
~          & H8S                & BE & ~      & s390       & S-390              & BE & ~     \\\hline 
hppa       & HP-PA              & BE & 32     & sparc      & SPARC              & BE & 32    \\\hline 
ia64       & IA-64              & LE & 128    & sparc64    &                    & BE & 32    \\\hline 
~          & IQ2000             & BE & ~      & ~          & Stormy16           & LE & ~     \\\hline 
~          & M32R               & BI & 16-32  & ~          & WASM               & LE & ~     \\\hline 
m68k       & M68K               & BE & ~      & amd64      & X86-64             & LE & 8-120 \\\hline 
~          & M88K               & BI & 32     & i386       & X86                & LE & 8-120 \\\hline 
~          & MCore              & BE & 16     & ~          & Xtensa             & BI & 16-24 \\\hline 
mips64el   & ~                  & LE & 32     & ~          & Z80                & LE & 8-32  \\\hline
~          & MIPS16             & BI & 16     & x32        & ~                  & LE & ~     \\\hline 
mips       & MIPSeb             & BE & 32     & powerpc    & ~                  & BE & 32    \\\hline 
mipsel     & MIPSel             & LE & 32     & powerpcspe & ~                  & BE & 32    \\\hline 
\end{tabular}
\end{table}

\noindent
CpuRec is used to detect fixed/variable instruction size and fixed instruction size, as it is more balanced across the classes these ISA characteristics contain than the IsaDetect datasets. However, CpuRec is still unbalanced for the fixed instruction size feature. Fixed instruction sizes tend to be 128 bits or less, and common sizes can be but are not limited to 8-, 16-, 24-, 32-, or 64-bits. This range of potential sizes makes it difficult to create balanced datasets, as some fixed instruction sizes will be less common than others.
\\\\
Experiments are conducted with the machine learning library \texttt{scikit-learn}. The selected classifiers are chosen to match those based on the paper from Kairajärvi et al. \cite{isadetect}. Using multiple classifiers allows for a more comprehensive evaluation of the data, as different models may capture different aspects of the underlying patterns.
\\\\
Choosing hyperparameters for classifiers and parameters for engineered features is essential to producing well-performing models. These values are found using a grid search for the classifiers to which this applies (LogisticRegression and Support Vector Classifier). Following the approach from Kairajärvi et al. \cite{isadetect}, various powers of ten are used to find parameters for classifiers. Powers of two are used to find the parameters for the engineered features, as the classes of the targeted ISA characteristics commonly are powers of two.
\\\\
Model performance is evaluated by calculating the model accuracy, which is defined below in Equation \ref{eq:model_accuracy}.
\begin{equation}\label{eq:model_accuracy}
     model\_accuracy = \frac{correct\_classifications}{total\_classifications}
\end{equation}
The model accuracy from Equation \ref{eq:model_accuracy} above is used with every model created with LOGOCV to calculate the accuracy of a feature. Feature accuracy depends on the hyperparameters, specific classifier, and dataset. The equation for feature accuracy is shown below in Equation \ref{eq:feature_accuracy}.
\begin{equation}\label{eq:feature_accuracy}
    feature\_accuracy = \sum_{n=1}^{m}\frac{model\_accuracy_n}{m},
\end{equation}
where $m$ is the number of ISAs (groups) used in the LOGOCV.
\\\\
Feature accuracy is measured in comparison against a baseline. Generally, features performing better than the baseline contain data that is helpful in classifying the targeted ISA characteristic. These features perform better than a random guess, allowing them to assist in documenting the ISA. The baseline is defined as the ratio resulting from classifying all samples as the most frequent class. The formula for the baseline is shown below in Equation \ref{new:eq:baseline}.

\begin{equation}\label{new:eq:baseline}
baseline = \frac{most\ frequent\ class\ count}{all\ count}
\end{equation}

\noindent
Detection of endianness focuses on differentiating binary files of big and little endianness. 
Two features are used for endianness detection, both inspired by previous research. The first of these is the Bigrams feature, which contains the frequency of every bigram from \texttt{0x0000} to \texttt{0xffff}. The second is the EndiannessSignatures feature, which includes the four selected bigrams from Clemens' \cite{automatic_classification_of_object_code_using_machine_learning} study (see Section \ref{new:20:background}): \texttt{0xfffe}, \texttt{0xfeff}, \texttt{0x0001} and \texttt{0x0100}. Due to the large dimensionality resulting from the full bigrams feature ($256^2$), only 100 binary files are used per ISA from the IsaDetectFull dataset when training and testing with this feature. This is to reduce the computational cost of training at the expense of the risk of overfitting data due to the curse of dimensionality. In practice this should not be a problem because the LOGOCV ensures an entire ISA is left out and used for testing. In contrast, all binary files are used with the EndiannessSignatures feature.
\\\\
A feature using autocorrelation has been engineered to detect fixed/variable instruction size and fixed instruction size. This feature is named AutoCorrelation and is defined in Equation \ref{eq:autocorr1} to \ref{eq:autocorr} below.
\\\\
\noindent
The computation of the AutoCorrelation feature is based on the Pearson correlation $r(x, y)$, defined in Equation \ref{eq:autocorr1}:

\begin{equation}\label{eq:autocorr1}
    r(x, y) = \frac{n\sum xy - (\sum x)(\sum y)}{\sqrt{[n\sum x^2 - (\sum x)^2][n\sum y^2 - (\sum y)^2]}}, 
\end{equation}
where $x$ and $y$ are the variables (original and lag windows) and $n$ refers to the number of samples in $x$ and $y$. The autocorrelation function is computed using the \texttt{pandas.autocorr} function from the \texttt{pandas} library in Python. Its mathematical formulation is shown in Equation \ref{eq:autocorr2}, where $s$ represents the series of bytes in a given binary file, and $k$ represents the given lag:

\begin{equation}\label{eq:autocorr2}
    f(k) = r(\{s_i\ |\ 1 \leq i \leq (|s| - k)\}, \{s_j\ |\ k \leq j \leq |s|\})
\end{equation}

\noindent The autocorrelation values, $f(k)$, are used to calculate the AutoCorrelation feature, as shown in Equation \ref{eq:autocorr}, where $l$ is the lag parameter of the AutoCorrelation feature, corresponding to the max lag used when calculating the autocorrelation values $f(k)$:

\begin{equation}\label{eq:autocorr}
    \text{AutoCorrelation} = \{ f(k)\ |\ 1 \leq k \leq l \}
\end{equation}

\noindent
The AutoCorrelation feature shown above in Equation \ref{eq:autocorr} calculates autocorrelation values by calculating the Pearson correlation $r(x, y)$ for a binary file and a specified range of lagged versions of itself. The aim is to discover periodicity resulting from repetitions of full or partial instructions at regular intervals corresponding to a multiple of the underlying fixed instruction size. In other words, we would expect autocorrelation peaks for lags that are an integer multiple of the fixed instruction size. For ISAs with fixed instruction sizes, this would result in a series with a general periodicity equal to the instruction size in bytes. The same would not be true for ISAs with variable instruction sizes, allowing for discrimination between these two classes.
\\\\
Specifically, the AutoCorrelation feature is first used to determine whether the ISA has a fixed or variable instruction size. If the ISA is classified as having fixed-size instructions, the AutoCorrelation feature is then used to determine the specific instruction size by analyzing the periodicity in the autocorrelation values.


\newcommand{\resultfiglabel}[2]{fig:research_#1_#2}

\section{Results}\label{new:40:results}

\subsection{Experimental Setup}
All experiments were run on an AMD EPYC 7742 64-Core server with 128GB RAM running Debian Linux 11 (bullseye) and Linux kernel 5.10.0-22-amd64. Versions of the software libraries were Python 3.9.2, scikit-learn 1.3.2, SciPy 1.6.0, and pandas 1.1.5. The source code used in the experiments is publicly available on GitHub\footnote{\url{https://github.com/joffe97/isa\_detection} (commit de355e0)}.



\noindent\\
\textit{LogisticRegression} and \textit{Support Vector Classifier} (\textit{SVC}) have a regularization parameter $C$ which we tuned via grid search. The results of the hyperparameter tuning are presented in Table \ref{new:tab:hyperparams}.

\newcommand{\hyperparamrowvalues}[2]{$10$\textsuperscript{$\wedge$}${#1}$ & $10$\textsuperscript{$\wedge$}${#2}$}
\newcommand{\hyperparammultirow}[3]{\multirow{#1}{#2}{{#3}}}
\renewcommand{\arraystretch}{1.1}
\begin{table}[H]
    \centering
    \caption{Hyperparameter values of $C$ for LogisticRegression (LR) and Support Vector Classifier (SVC) discovered via grid search.}
    \begin{tabular}{|c|c||c|c|}
        \hline
        \textbf{ISA characteristic}                        & \textbf{Data feature}        & \textbf{LR $C$}  & \textbf{SVC $C$}   \\\hline
        \hyperparammultirow{2}{*}{Endianness}       & EndiannessSignatures  & \hyperparamrowvalues{10}{11}                  \\\cline{2-4}
                                                    & Bigrams               & \hyperparamrowvalues{5}{3}                    \\\hline
        Fixed/variable instruction size             & AutoCorrelation       & \hyperparamrowvalues{0}{0}                    \\\hline
        Fixed instruction size                      & AutoCorrelation       & \hyperparamrowvalues{1}{1}                    \\\hline
    \end{tabular}
    \label{new:tab:hyperparams}
\end{table}
\renewcommand{\arraystretch}{1.0}

\noindent
The tuned lag parameters of the AutoCorrelation feature for every configuration of classifier and ISA characteristic are shown below in Table \ref{new:tab:tuning_feature_instsize_autocorr}. Each row in the table consists of lags tuned for configurations of classifiers and ISA characteristics, where the first column is the classifier, and the second and third columns show the tuned lags of the ISA characteristic. 

\newcommand{\parametertuningfeature}[3]{Tuned parameters for the \texttt{{#1}} feature, targeting {#2}. Models were trained and tested on the {#3} dataset.}
\begin{table}[H]
    \centering
    \caption{Tuned lag parameters for the AutoCorrelation feature.}
    \begin{tabular}{|c||c|c|}
\hline
\multirow{2}{*}{\textbf{Classifier}} & \textbf{Lag for} & \textbf{Lag for} \\
                                     & \textbf{fixed/variable instruction size} & \textbf{fixed instruction size} \\\hline
1NeighborsClassifier                 & 256  & 32  \\\hline
3NeighborsClassifier                 & 256  & 128 \\\hline
5NeighborsClassifier                 & 512  & 512 \\\hline
DecisionTreeClassifier               & 128  & 128 \\\hline
GaussianNB                           & 32   & 256 \\\hline
LogisticRegression                   & 128  & 128 \\\hline
MLPClassifier                        & 1024 & 32  \\\hline
RandomForestClassifier               & 256  & 256 \\\hline
SVC                                  & 128  & 64  \\\hline
    \end{tabular}
    \label{new:tab:tuning_feature_instsize_autocorr}
\end{table}

\noindent
The baselines for the three targeted ISA characteristics are shown below in Table \ref{new:tab:baselines}. Note that the baselines for the EndiannessSignatures and Bigrams features targeting endianness differ; Bigrams uses the same number of files per ISA (100), in contrast to EndiannessSignatures (full dataset). The first two columns contain ISA characteristics and classifier data features, respectively, where each row represents a configuration used in the experiments, while the third column contains the baseline result accuracy for each configuration.

\begin{table}[H]
    \centering
    \caption{Feature baselines.}
    \begin{tabular}{|c|c||c|c|}
        \hline
        \textbf{ISA characteristic}                        & \textbf{Data feature}      & \textbf{Baseline accuracy} \\\hline
        \hyperparammultirow{2}{*}{Endianness}       & EndiannessSignatures  & 0.556             \\\cline{2-3}
                                                    & Bigrams               & 0.545             \\\hline
        Fixed/variable instruction size             & AutoCorrelation       & 0.581             \\\hline
        Fixed instruction size                      & AutoCorrelation       & 0.680             \\\hline
    \end{tabular}
    \label{new:tab:baselines}
\end{table}




\newcommand{\includegraphicsresultfig}[1]{\includegraphics[width=0.93\linewidth,trim={0.0cm 0.2cm -1.2cm 0.7cm},clip]{#1}}

\newcommand{\resultfig}[4]{
    \begin{figure}[H]
        \centering
        \includegraphics[width=0.8\linewidth,trim={0.2cm #4 0.7cm 2.4cm},clip]{Images_paper/#1_#2.eps}
        \caption{#3}
        \label{\resultfiglabel{#1}{#2}}
    \end{figure}
}

\subsection{Experiments}
\subsubsection{Endianness} results are presented below in Figure \ref{fig:new:result_endianness}. As mentioned in Section \ref{new:30:methodology}, these models are trained on the IsaDetectFull dataset, where EndiannessSignatures uses all files, and Bigrams uses 100 per ISA. 
\\\\
The results show that the EndiannessSignatures feature generally performs better than the Bigrams feature when detecting endianness. All models perform better than the baselines presented in Table \ref{new:tab:baselines} of $0.556$ for EndiannessSignatures and $0.545$ for Bigrams. The best-performing models for EndiannessSignatures and Bigrams achieve an accuracy of $0.994$ and $0.986$, respectively.
\\\\
\begin{figure}[H]
    \centering
    \includegraphicsresultfig{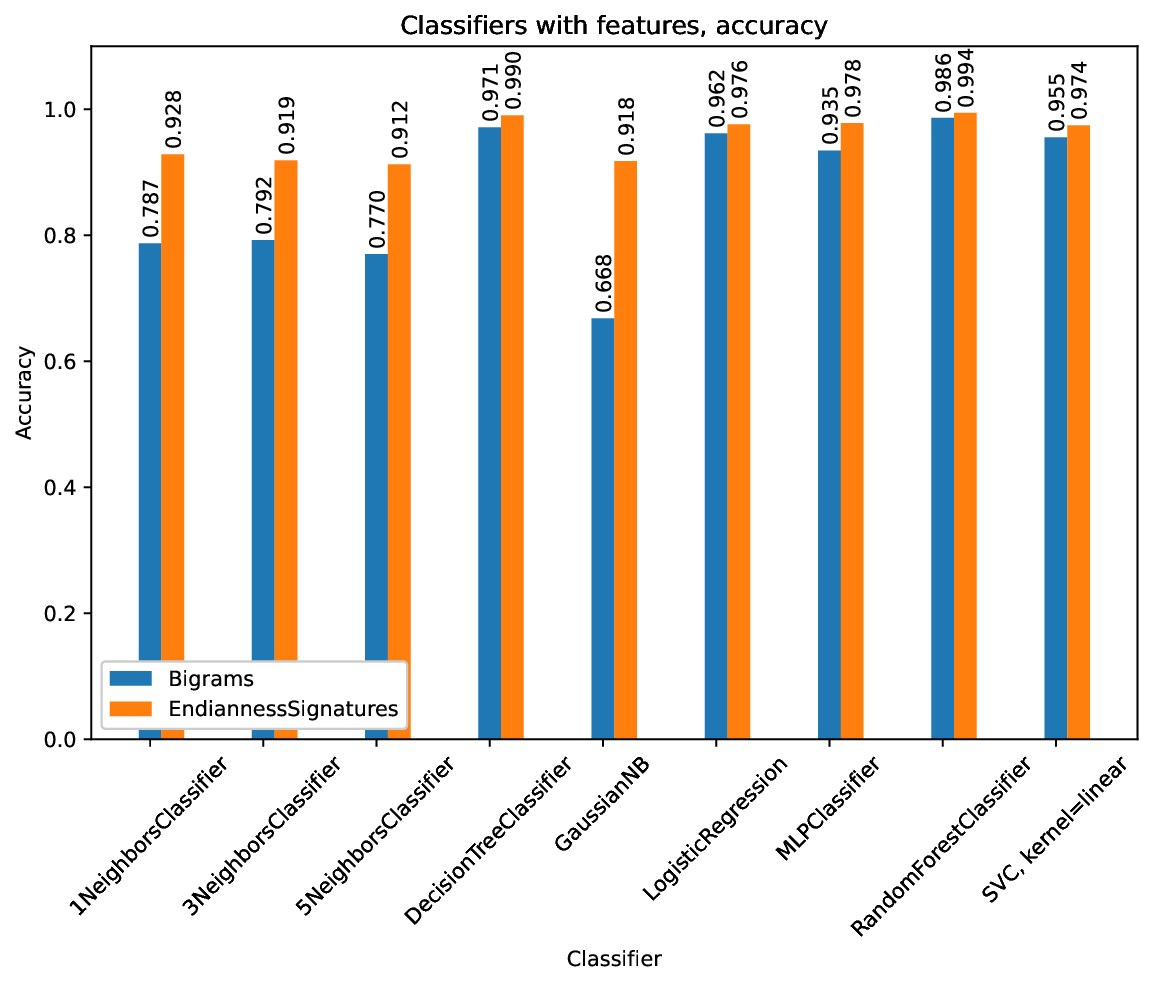}
    \caption{EndiannessSignatures and Bigrams models targeting endianness (IsaDetectFull dataset).}
    \label{fig:new:result_endianness}
\end{figure}

\subsubsection{Fixed/variable instruction size} results are presented below in Figure \ref{\resultfiglabel{CpuRec}{typemean}}. The plot in this figure is generated by calculating the average accuracy of the ISAs from the CpuRec dataset, grouped by fixed and variable instruction sizes. Lags in the x-axis represent numbers of bytes.
\\\\
The plot shows that the AutoCorrelation feature can differentiate between ISAs of fixed and variable instruction sizes, as the plotted values show a clear difference between the two classes. ISAs of fixed instruction sizes are shown to have greater peaks than ISAs of variable instruction sizes at every fourth byte lag. This observed difference indicates that the AutoCorrelation feature is suitable for differentiation between fixed and variable instruction sizes.
\\\\
\resultfig{CpuRec}{typemean}{AutoCorrelation mean values for fixed and variable instruction sizes  (CpuRec dataset).}{3.1cm}

\noindent
Figure \ref{fig:new:result_isvar} below shows that the AutoCorrelation feature can classify fixed/variable instruction size with an accuracy of $0.860$, which is significantly greater than the baseline of $0.581$.
\begin{figure}[H]
    \centering
    \includegraphicsresultfig{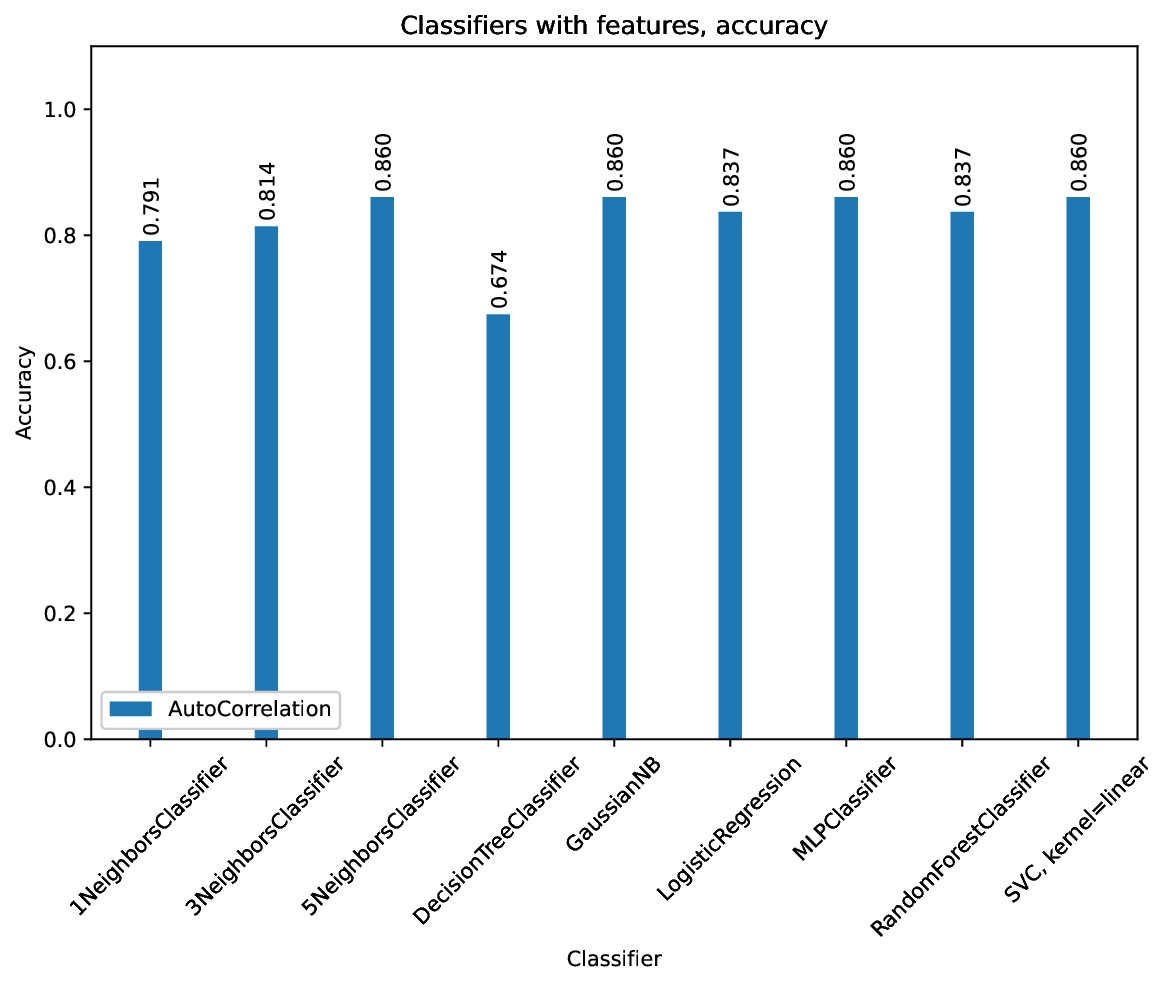}
    \caption{AutoCorrelation models targeting fixed/variable instruction size (CpuRec dataset).}
    \label{fig:new:result_isvar}
\end{figure}

\subsubsection{Fixed instruction size} results are presented below in Figure \ref{\resultfiglabel{CpuRec}{sizemean}}. The values in this plot are generated similarly to the plot for fixed/variable instruction size in Figure \ref{\resultfiglabel{CpuRec}{typemean}}. The average values calculated from the AutoCorrelation feature are grouped and plotted for a range of lags.
\\\\
The plot shows there are generally peaks on lags equal to an integer multiple of classes' instruction sizes in bytes. This implies that the AutoCorrelation feature might also be suitable for fixed instruction size detection. The exception to this pattern is the data for 24-bit instruction size. It should be noted that only one file from a single ISA in the CpuRec dataset has an instruction size of 24 bits. This means the irregularity could result from a unique property of the specific binary file or ISA unrelated to instruction size.
\\\\
\resultfig{CpuRec}{sizemean}{AutoCorrelation mean values for fixed instruction sizes (CpuRec dataset).}{3.1cm}

\noindent
Accuracies of the classifiers using the AutoCorrelation features in fixed instruction size detection are shown below in Figure \ref{fig:new:result_instsize}. This Figure shows that the models can detect fixed instruction size with an accuracy of $0.880$. This is significantly greater than the baseline of $0.680$, meaning that the AutoCorrelation feature is suitable for classifying fixed instruction size. 
\\\\
It should be noted that the CpuRec dataset contains only one ISA each of 24 and 128 bits. Due to the use of LOGOCV, fixed instruction sizes belonging to only one ISA will not be able to be detected when generating results. Specifically, models testing one of these ISAs will not be trained on any instruction size of the same ISA, leading to the models being unable to classify it. This means that 2 of 77 ISAs will always be classified incorrectly. This should be considered in the analysis of resulting accuracies when targeting fixed instruction size.
\begin{figure}[H]
    \centering
    \includegraphicsresultfig{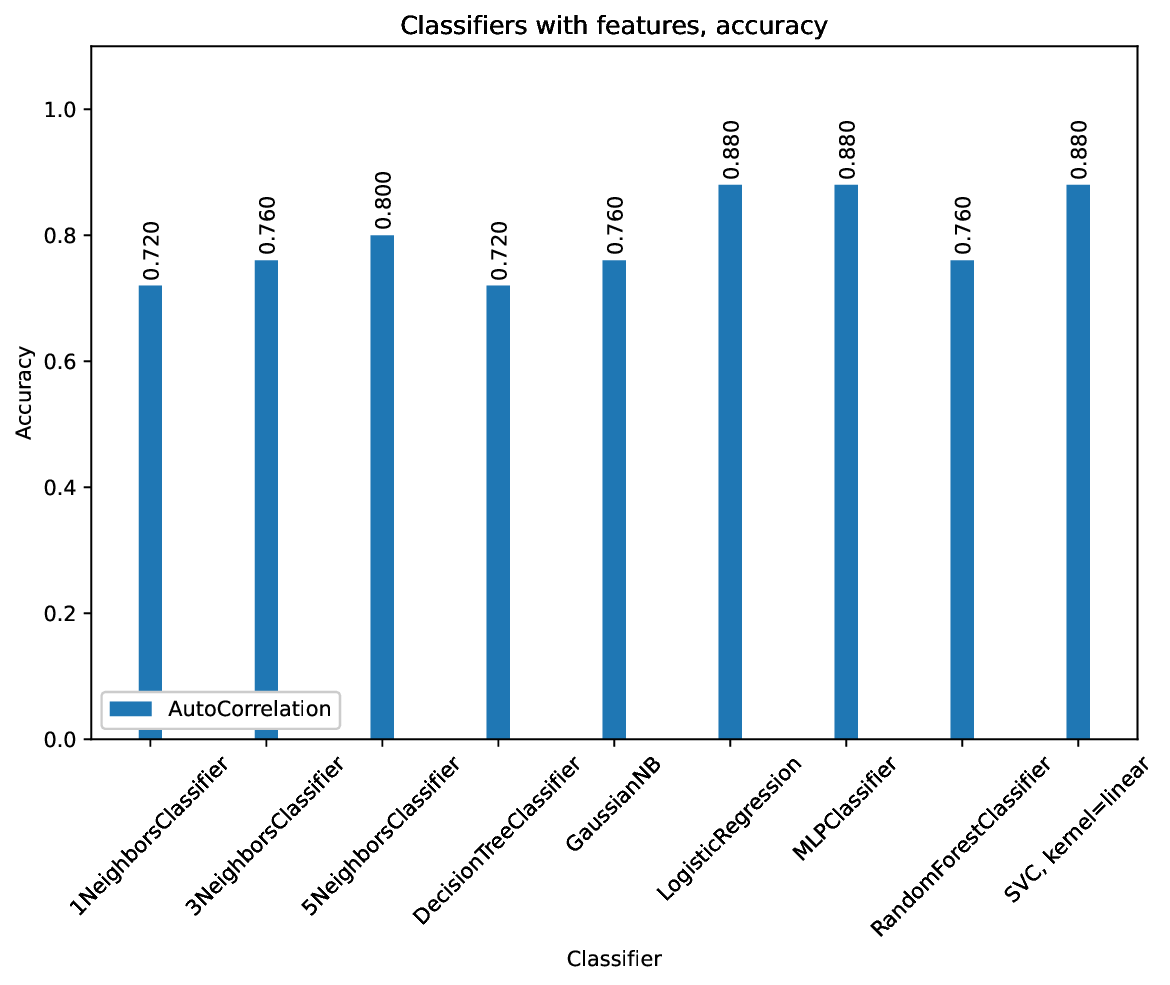}
    \caption{AutoCorrelation models targeting fixed instruction size  (CpuRec dataset).}
    \label{fig:new:result_instsize}
\end{figure}
\section{Discussion}\label{new:50:discussion}
Regarding RQ1, the results from Section \ref{new:40:results} show that machine learning can reliably detect ISA characteristics from binary programs by using features containing information related to the targeted ISA characteristics. The results show this to be true for the ISA characteristics targeted in this research, showing that models consistently achieve accuracies well above the baseline. The experiments show that the classifiers' accuracies vary based on the targeted ISA characteristic. For example, classifiers that perform well in detecting endianness, such as the Decision Tree Classifier, do not necessarily achieve high accuracy in detecting fixed/variable instruction size or fixed instruction size. This difference in performance highlights the benefit of running experiments with multiple classifiers.
\\\\
Answering RQ2 requires analyzing the results presented in Section \ref{new:40:results} to determine whether the ISA characteristics are detected reliably across a wide range of ISAs. The results show that we are able to detect endianness, fixed/variable instruction size, and fixed instruction size consistently with an accuracy greater than the baseline. Since the experiments use datasets with broad ranges of ISAs, we can conclude that the proposed approaches do lead to reliable detection of ISA characteristics across a wide range of ISAs.
\\\\
The results from endianness detection show that bigrams can detect endianness from binary programs with any arbitrary ISA, with a greater accuracy than the baseline. The experiments also demonstrate that the four bigrams included in the EndiannessSignatures feature further improve the accuracy, inferring that excluding non-descriptive bigrams increases accuracy. Although the curse of dimensionality cannot be ruled out as a factor, LOGOCV ensures that any overfitting means the results are conservative (tending towards poorer accuracy than better because the target ISA has not been seen during training).
\\\\
The experiments show that features using autocorrelation can differentiate ISAs of fixed and variable instruction sizes and classify specific fixed instruction sizes with greater accuracy than the baseline. This demonstrates that signal processing is applicable in the field of reverse engineering for ISA classification.
\\\\
If we presume that correctly classifying a binary program into one of a set of known ISAs also yields the endianness or instruction size, then our results are comparable to those of Kairajärvi et al. \cite{isadetect}. Kairajärvi et al. can correctly assign a known ISA to a binary program with an accuracy of $98\%$. The results from this paper show that the proposed methods are not able to retrieve ISA characteristics with the same accuracy, as the chance of false classifications accumulates for every ISA characteristic detection. This shows that the methods proposed by Kairajärvi et al. are more suited for detecting ISA characteristics when the ISA is known. As mentioned, the method of Kairajärvi et al. cannot detect ISA characteristics for unknown ISAs. The results presented in Section \ref{new:40:results} show that the methods utilized in this paper can do so with great accuracy. These methods are, therefore, more suitable for detecting ISA characteristics where the ISA is unknown.




\section{Conclusion}\label{new:60:conclusion}
This work presents various features and methods for detecting endianness and instruction size characteristics in binary programs with unknown ISAs. Experiments with features using bigrams are performed, demonstrating that they are well suited for endianness detection of binary programs with unknown ISAs. Fixed/variable instruction size and fixed instruction size are also shown to be accurately predicted for such binary programs using a feature that use autocorrelation.
\\\\
Conducting experiments using a range of classifiers from the \texttt{scikit-learn} library in Python has proven valuable, as results demonstrate that classifier choice significantly impacts accuracy, enabling the application of models best suited for specific classifications.
\\\\
The use of Leave-One-Group-Out Cross-Validation trains a model without any direct knowledge of the targeted binary program's ISA. This ensures experiments that simulate the detection of ISA characteristics for truly unknown ISAs.
\\\\
Several possibilities exist for future work. One involves creating a more extensive balanced dataset, which would lead to more robust experiments. More training data also allows models to discover more detailed patterns related to the various target classes. Balancing this dataset leads to models being more equally influenced by all classes instead of favoring their ability to detect the most frequently occurring class.
\\\\
The disadvantage of unbalanced datasets could be mitigated for code-only binary files by splitting them into smaller chunks and training the models on an equal number of chunks from each target class. Future work could explore whether this approach improves accuracy for under-represented classes. However, care should be taken to avoid splitting instructions at non-boundary locations, which could negatively impact results.
\\\\
Extending the research by developing additional methods for discovering other ISA characteristics would further streamline the reverse engineering of binary files with unknown or undocumented ISAs. Suggested ISA characteristics include word size, register count, instruction format, opcode encoding, and subroutine boundaries (e.g., isolation of CALL/RET opcodes as in \cite{pettersen2024graphdiscoverybinaryprograms}).
 

\bibliography{references}

\begin{thebibliography}{1}

\bibitem{arm_glossary_isa}
{Arm}.
\newblock Glossary - instruction set architecture (isa), 2024.

\bibitem{automatic_classification_of_object_code_using_machine_learning}
John Clemens.
\newblock Automatic classification of object code using machine learning, 2015.

\bibitem{cpu_rec}
Louis Granboulan.
\newblock cpu\_rec, 6 2024.

\bibitem{isadetect}
Sami Kairaj\"{a}rvi, Andrei Costin, and Timo H\"{a}m\"{a}l\"{a}inen.
\newblock Isadetect: Usable automated detection of cpu architecture and endianness for executable binary files and object code, 2020.

\bibitem{towards_usable_cpu_architecture_detection}
Sami Kairajärvi, Andrei Costin, and Timo Hämäläinen.
\newblock Towards usable automated detection of cpu architecture and endianness for arbitrary binary files and object code sequences, 2019.

\bibitem{elisa_eliciting_isa_of_raw_binaries_for_fine_grained_code_and_data_separation}
Pietro~De Nicolao, Marcello Pogliani, Mario Polino, Michele Carminati, Davide Quarta, and Stefano Zanero.
\newblock Elisa: Eliciting isa of raw binaries for fine-grained code and data separation, 2018.

\bibitem{pettersen2024graphdiscoverybinaryprograms}
Håvard Pettersen and Donn Morrison.
\newblock Call graph discovery in binary programs from unknown instruction set architectures, 2024.

\bibitem{natural_language_processing_approach_for_isa_identification}
Dinuka Sahabandu, Sukarno Mertoguno, and Radha Poovendran.
\newblock A natural language processing approach for instruction set architecture identification, 2022.

\bibitem{tis_elf}
{TIS Committee}.
\newblock Executable and linking format (elf), 1995.

\end{thebibliography}

%
%

%
%




\end{document}